\documentclass[a4paper]{amsart}
\usepackage[T1]{fontenc}
\usepackage[utf8]{inputenc}
\usepackage{graphicx}

\usepackage{amsmath,amssymb,amsthm,amsfonts,enumerate,url,fancyhdr,mathbbol,mathrsfs,color}
\usepackage{hyperref}

\usepackage[utf8]{inputenc}
\usepackage{pdfsync}
\def\me{\mathsf{e}}
\def\mv{\mathsf{v}}
\def\:{\thinspace:\thinspace}
\newtheorem{theo}{Theorem}
\newtheorem{defi}[theo]{Definition}

\theoremstyle{definition}

  \def\mG{\mathsf{G}}
  \def\mH{\mathsf{H}}
  \def\mV{\mathsf{V}}
  \def\mE{\mathsf{E}}

\def\mv{\mathsf{v}}
 \def\me{\mathsf{e}}

  \def\mf{\mathsf{f}}

\thanks{
This research has been supported by the Land Baden--W\"urttemberg in the framework of the \emph{Juniorprofessorenprogramm} -- research project on ``Symmetry methods in quantum graphs'' -- and by the Center for Interdisciplinary Research (ZiF) in Bielefeld in the framework of the cooperation group on ``Discrete and continuous models in the theory of networks".
}
\begin{document}

\title[Laplacians on quantum hypergraphs]{Laplacians on quantum hypergraphs}

\author{Delio Mugnolo}%

\address{Institut f\"ur Analysis, Universit\"at Ulm, 89069 Ulm, Germany}
\email{delio.mugnolo@uni-ulm.de}

\begin{abstract}
We introduce quantum hypergraphs, in analogy with the theory of quantum graphs developed over the last 15 years by many authors. We emphasize some problems that arise when one tries to define a Laplacian on a hypergraph.
\end{abstract}
\maketitle                   


A hypergraph is a pair $\mH:=(\mV,\mE)$, where $\mV$ is a set of \emph{nodes} but, unlike in the more usual case of a \emph{graph}, $\mE$ is some set each of whose elements $\me$ -- so-called \emph{hyperedges} -- is an unordered $n_\me$-tuple of elements of $\mV$, where each $n_\me\in \mathbb N$ may depend on $\me$. Thus we obtain graphs as a special case of hypergraphs if $n_\me\equiv 2$. Conversely, a graph is canonically associated with $\mH$ upon replacing each hyperedge $\me$ by a collection of $\frac12 n_\me(n_\me-1)$ edges that connect any two distinct elements of $\me$: This graph is called the \emph{section} of $\mH$ in~\cite{Ber73}. Observe that the connectivity of a hypergraph cannot be reconstructed from its section.

If an orientation is assigned to each edge of a simple graph $\mG:=(\mV,\mE)$, i.e., if each edge is regarded as an \emph{ordered} pair $\me\equiv(\me_{\rm init},\me_{\rm term})\in \mV\times \mV\setminus\{(\mv,\mv):\mv\in \mV\}$, then it is possible to define the \emph{incidence matrix} $\mathcal I=(\iota_{\mv\me})$ by 
$${\iota}_{\mv \me}:=\left\{
\begin{array}{ll}
+1 & \hbox{if}~\mv= \me_{\rm init},\\
-1 & \hbox{if}~\mv= \me_{\rm term},\\
0 & \hbox{otherwise},
\end{array}\right.$$ 
and the \emph{graph Laplacian} by $\mathcal L:=\mathcal I\mathcal I^T$. Of course, $\mathcal L$ is a symmetric and positive semidefinite operator on the \emph{node space} $\mathbb R^\mV$.

The first attempts to develop a systematic theory of operators on $\mathbb R^\mV$ -- and in particular of graph Laplacians -- date back to the 1970s: This \emph{spectral graph theory} is nowadays rather mature, cf.~\cite{GodRoy01,BroHae12}. One may want to introduce a hypergraph Laplacian as the graph Laplacian of the non-oriented hypergraph's section. This choice is made e.g.\ in~\cite{ZhoHuaSch07} and a few subsequent papers on spectral clustering, but we argue that in this way the essential non-binary structure of a hypergraph is lost. 

An alternative approach seems to be more accurate and ultimately more appropriate for our purposes: Oriented hypergraphs have been introduced independently by many authors, see e.g.\ the historical remarks in~\cite{GalLonPal93}, but  it was only in the last few years that an algebraic \emph{hyper}graph theory has been developed by their means. Indeed, upon regarding each hyperedge as an oriented pair $\me\equiv(\me_{\rm init},\me_{\rm term})$ of disjoint subsets of $\mV$ it is possible to introduce an incidence matrix $\mathcal I=(\iota_{\mv\me})$ by
$${\iota}_{\mv \me}:=\left\{
\begin{array}{ll}
+1 & \hbox{if}~\mv\in \me_{\rm init},\\
-1 & \hbox{if}~\mv\in \me_{\rm term},\\
0 & \hbox{otherwise}.
\end{array}\right.$$ 
The algebraic properties of a hypergraph's incidence matrix are not as well understood as in the graph case. E.g., the combinatorial meaning of the rank and the co-rank of $\mathcal I$ seems to be unknown. Nevertheless, it is again possible to define a symmetric, positive semidefinite hypergraph Laplacian by $\mathcal L:=\mathcal I\mathcal I^T$. Its algebraic properties have been studied in~\cite{Chu93} and more recently in~\cite{RefRus12,Rus13,HorMugPol14}, while here we briefly dip into the time-continuous evolution equation 
\[
\frac{df}{dt}(t,\mv)=-\mathcal Lf(t,\mv),\qquad t\ge 0,\ \mv\in \mV\ .
\]

The associated Cauchy problem is well-posed if the hypergraph is finite, but in general it is not associated with a Markov process, as one sees already in the simple case of a hypergraph consisting of three nodes and one hyperedge $\me$ with $\me_{\rm int}=\{\mv_1,\mv_2\}$ and $\me_{\rm term}=\{\mv_3\}$. In this case the incidence matrix, the hypergraph Laplacian, and the generated semigroup are
\[
\mathcal I=\begin{pmatrix}
-1 \\ -1 \\ 1
\end{pmatrix},\quad \mathcal L=\begin{pmatrix}
1 & 1 & -1\\  1 & 1 & -1\\ -1 & -1 & 1
\end{pmatrix},\quad \hbox{ and }\quad
e^{-t\mathcal L}=
\frac{1}{3}\begin{pmatrix}
e^{3t}+2 & e^{3t}-1 & -e^{3t}+1\\  
e^{3t}-1 & e^{3t}+2 & -e^{3t}+1\\  
-e^{3t}+1 & -e^{3t}+1 & e^{3t}+2\\  
\end{pmatrix},\ t\ge 0\ .
\]

One can actually prove that if $\mH$ is a proper hypergraph (i.e., not a graph), then its Laplacian \emph{always} has both positive and negative off-diagonal entries (this also follows from~\cite{Chu93}) and hence cannot generate a positivity-preserving semigroup. 

In view of the attention devoted to quantum graphs since~\cite{KotSmi97}, cf.\ \cite{BerKuc13,Mug14}, it is natural to wonder whether a quantum hypergraph (and a Laplacian thereon) can also be defined. 
 The previous observation on $(e^{-t\mathcal L})_{t\ge 0}$ shows that switching from hypergraphs to \emph{quantum} hypergraphs, and in particular defining a differential operator whose discretization is $\mathcal L$, is less obvious than in the case of graphs. The sought-after differential operator cannot be a collection of Laplacians $\Delta_\me$ with local boundary conditions on polygons $\Omega_\me\subset \mathbb R^2$, $\me\in \mE$, as the common drawing of hypergraphs may naively suggest: Indeed, for such collections generate positivity-preserving semigroups, cf.~\cite{CarMug09}. 
Instances of Laplacian-type differential operators that generate a non-positivity-preserving semigroup are known, though: E.g., for an $\mE\times \mE$-matrix-valued function $C$ the operator
\[
f\mapsto \frac{d}{dx}\left( C\frac{df}{dx}\right)
\]
with some natural boundary conditions was studied in~\cite{CarMugNit08} -- remarkably, that investigation had a different motivation. It was proved in~\cite{CarMugNit08} that this operator is self-adjoint if $C(x)$ is positive definite (uniformly in $x$); but that the generated semigroup is positivity-preserving if and only if $C(x)$ is diagonal for all $x$. Hence we propose the following.

\begin{defi}\label{defi:qgraph}
Let $\mH=(\mV,\mE)$ be an oriented hypergraph such that $\me_{\rm init}\ne \emptyset\ne \me_{\rm term}$ for all $\me\in\mE$. We call \emph{oriented section} $\mG_\mH=(\mV,\mE_\mH)$ of an oriented hypergraph $\mH=(\mV,\mE)$ the oriented graph obtained replacing each oriented hyperedge $(\me_{\rm init},\me_{\rm term})\in \mE$ by a collection of $M_\me:=|\me_{\rm init}|\cdot|\me_{\rm term}|$ oriented edges that go from each element of $\me_{\rm init}$ to each element of $\me_{\rm term}$.

Let ${\bf M}:=|\mE_\mH|=\sum_{\me\in \mE}M_\me$ and let $C$ be the ${\bf M}\times {\bf M}$-diagonal-block matrix whose block indexed by $\me$ is the $M_\me\times M_\me$-matrix each of whose entries is the scalar $M_\me^{-1}$. Then the \emph{quantum hypergraph Laplacian on $\mH$} is 
\[
\Delta:(f_\mf)_{\mf\in \mE_\mH}\mapsto \left(\frac{d}{dx}\left( C\frac{df_\mf}{dx}\right)\right)_{\mf\in \mE_\mH}\ ,\qquad f\in H^2(0,1;\mathbb C^{\mE_\mH})\equiv H^2(0,1)\otimes \mathbb C^{\bf M}.
\]
\end{defi}

We have in particular identified each edge $\mf\in \mE_\mH$ of $\mG_\mH$ with an interval $[0,1]$. 
If in particular $\mH$ is actually a graph, i.e., $\mG_\mH\equiv\mH$, then $M_\me= 1$ for all $\me\in \mE$ and we recover the usual quantum graph Laplacian. 

The matrix $C$ in Definition~\ref{defi:qgraph} is symmetric -- indeed, it is an orthogonal projector. It is positive definite if and only if it is diagonal, i.e., if and only if $\mH$ is a graph. Thus, $\Delta$ cannot be proved to be self-adjoint on $L^2(\mG_\mH):=L^2(0,1)\otimes \mathbb C^{\bf M}$ by the results in~\cite{CarMugNit08}, regardless of the boundary conditions we impose. 
In order to overcome this difficulty, let us restrict $\Delta$ to
\[
\mathcal D:=\{ f\in H^1(\mG_\mH)\cap H^2(0,1)\otimes  \mathbb C^{\bf M}: f'_\mf(0)=f'_\mf(1)=0 \hbox{ for all }\me\in\mE \}\ ,
\]
where $H^1(\mG_\mH)$ is the space $f\in L^2(\mG_\mH)$ of those functions that are edgewise weakly differentiable and such that $Cf$ is continuous in the nodes.
A direct computation shows that $\Delta_{|\mathcal D}$ is a symmetric and positive semidefinite operator.
The associated form is not closed, but it is densely defined and hence by~\cite[Thm.~13.12]{Sch12} $\Delta_{|\mathcal D}$ admits self-adjoint extensions. If $\mH$ is a graph, then it has been proved in~\cite{Mug14c} that the usual quantum graph Laplacian -- i.e., the second derivative with continuity and Kirchhoff-type boundary conditions in the nodes -- agrees with the Friedrichs extension of $\Delta_{|\mathcal D}$. We regard the Friedrichs extension of $\Delta_{|\mathcal D}$ as the canonical Laplacian for general hypergraphs, too. We will describe its domain in a later paper.

\end{document}